# Two-dimensional Graphene Heterojunctions: the Tunable Mechanical Properties

Kang Xia[1], Haifei Zhan[1,2,3], and Yuantong Gu[*,1]

[1]School of Chemistry, Physics and Mechanical Engineering, Queensland University of Technology, Brisbane QLD 4001, Australia

[2]Institute of High Performance Computing, Agency for Science, Technology and Research, 1 Fusionopolis Way, Singapore 138632

[3]Department of Mechanical Engineering, National University of Singapore, 9 Engineering Drive 1, Singapore 117576

*Corresponding author Telephones: +61-7-31381009
E-mail address: yuantong.gu@qut.edu.au

**Abstract**

We report the mechanical properties of different two-dimensional carbon heterojunctions (HJs) made from graphene and various stable graphene allotropes, including α-, β-, γ- and 6612-graphyne (GY), and graphdiyne (GDY). It is found that all HJs exhibit a brittle behaviour except the one with α-GY, which however shows a hardening process due to the formation of triple carbon rings. Such hardening process has greatly deferred the failure of the structure. The yielding of the HJs is usually initiated at the interface between graphene and graphene allotropes, and monoatomic carbon rings are normally formed after yielding. By varying the locations of graphene (either in the middle or at the two ends of the HJs), similar mechanical properties have been obtained, suggesting insignificant impacts from location of graphene allotropes. Whereas, changing the types and percentages of the graphene allotropes, the HJs exhibit vastly different mechanical properties. In general, with the increasing graphene percentage, the yield strain decreases and the effective Young's modulus increases. Meanwhile, the yield stress appears irrelevant with the graphene percentage. This study provides a fundamental understanding of the tensile properties of the heterojunctions that are crucial for the design and engineering of their mechanical properties, in order to facilitate their emerging future applications in nanoscale devices, such as flexible/stretchable electronics.

Keywords: graphene, heterojunctions, graphyne, mechanical properties, molecular dynamics simulations

1. Introduction

Owing to its versatile flexibility, carbon is able to form three different hybridization states ($sp^1$, $sp^2$, $sp^3$) and various allotropes, such as fullerene, [1] carbon nanotube [2] and graphene [3]. Recent years, extensive researches have been conducted to explore the unique properties and potential applications of other 2D carbon networks, such as graphyne and graphdiyne [4, 5]. Like graphene (G), graphyne (GY) and graphdiyne (GDY) are also one-atom-thick sheet of carbon atoms but with different content of $sp^2$ hybridized bonds. The presence of the $sp^2$ carbon atoms destroys the regular hexagonal crystal lattice of the graphene, which allows for the formation of various types of GYs with different geometries. These GYs differ from each other with regard to the percentage of the acetylenic linkages (-C≡C-) in their structures. The percentages of the acetylenic linkages are 100%, 66.67%, 33.33% and 41.67% for α-, β-, γ-, and 6612-GYs, respectively. It is expected that the introduction of different densities of the linkages in GYs should make their mechanical properties distinctly different from those of graphene [6].

Several works have been devoted to investigate the electrical and thermal properties of those 2D graphene allotropes. It is reported that GY and GDY are semiconductors with direct transitions at the M and Γ points of the Brillouin zone [7-10]. Recently, considering the exceptional electronic mobility of graphene and the non-zero band gap of graphene or GDY, researchers proposed a heterojunctions (HJs) of graphene and GY or GDY. Such HJs based transistors are predicted to exhibit excellent switching behaviours without the severe contact resistance as in typical metal electrodes [10]. In addition, carbon allotropes own remarkable thermal properties. The thermal conductivity at room temperature is reported range from ~0.01 W/mK in amorphous carbon to above 2000 W/mK in diamond or graphene [11]. Recent studies show that, the thermal conductivity for GY geometries is observed to decrease monotonously with increasing number of acetylenic linkages between adjacent hexagons. Incorporating the percentage, type and distribution pattern of the GY or graphyne-like structures, the HJs are found to exhibit tuneable thermal transport properties [12]. Recent studies suggest that varying the γ-GY percentage will lead to a change in ON/OFF ratio on the order of $10^2$-$10^3$ [13], the GDY could server as the reducing agent and stabilizer for electroless deposition of highly dispersed Pd nanoparticles [14] and GDY could also enhance the power conversion efficiency of the organic-inorganic perovskite solar cells [15].

While engineering the structures of the graphene allotropes (to optimal their electrical or thermal properties), a substantial impact to their mechanical properties is also expected. However, a comprehensive understanding of how their mechanical performance will change with varying structures are still lacking of investigation, which is critical for their implementation and operation in the nanoscale devices. Therefore, in this work, the large-scale molecular dynamics (MD) simulations will be employed to probe the mechanical properties of graphene HJs. By taking the tensile deformation as a representative loading scenario, an in-depth discussion will be carried out to discuss how different graphene HJs will behave. Results obtained from this study are expected to guide the design of different HJs and eventually benefit their diversity engineering applications.

## 2. Computational details

To access how the HJs will perform while containing different percentages of graphene allotrope and graphene, a series of large-scale MD simulations were performed using the open-source LAMMPS code [16]. Two groups of samples have been considered, with the graphene allotrope either located in the middle or at the two ends of structure as illustrated in Figure 1, different percentages (ranging from 10% to 80%) of varies graphene allotropes (including α-, β-, γ- and 6612-GY, and GDY) have been considered. The percentage is simply the ratio between the area of the allotrope and the whole structure. All samples are symmetrical along the mid of the length direction and possess a similar initial size, i.e., $20.0 \times 4.0$ nm$^2$, with a quasi-armchair edge along the length direction. For discussion simplicity, the HJ is named according to the location and percentage of the graphene allotrope, e.g., the G-αGY20%-G indicates the sample has 20% of α-GY in the middle of the structure and 80% of graphene (denoted as G) at the two ends.

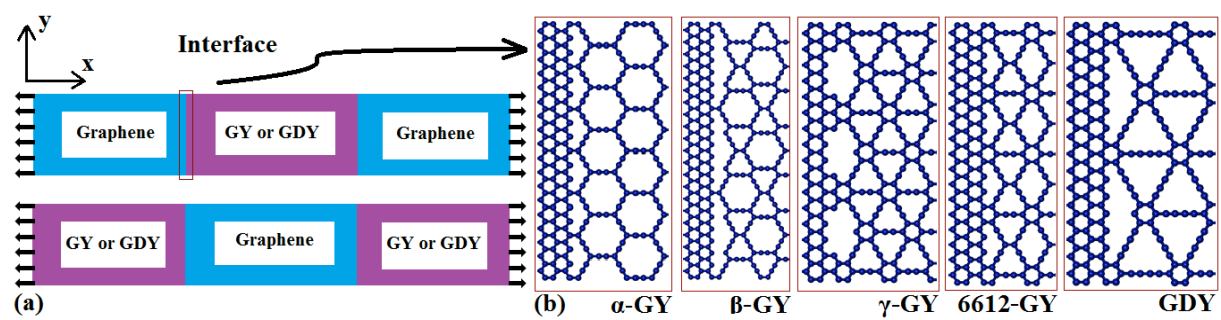

**Figure 1** (a) Schematic view of different models with graphene allotrope (GY/GDY) located either in the middle (upper figure) or at the two ends (lower figure). (b) Atomic representation of the interface between graphene and GY/GDY.

The popularly applied reactive empirical bond order (REBO) potential [17] was adopted to describe the interactions of carbon atoms, which has been shown to well represent the binding energy and elastic properties of graphene and graphene allotropes [18]. During the simulation, the HJs are firstly relaxed to a minimum energy state using the conjugate gradient algorithm. We then used the Nose-Hoover thermostat [19, 20] to equilibrate the HJs under the NVT ensemble for 500 ps at a time step of 1 fs. Finally, a constant velocity of 0.001 Å/ps was applied to one end of the HJs along the *x*-axis direction (Figure 1, corresponding to a strain rate $\sim 2.5 \times 10^{-8}$ ps$^{-1}$), while holding another end fixed. The equations of motion are integrated with time using a Velocity Verlet algorithm [21]. No periodic boundary conditions have been applied. The system temperature was maintained at 1 K during the simulation to minimize the influence from thermal fluctuations.

In order to overcome spuriously high tensile force when the carbon-carbon bond length is greater than 1.7 Å, the onset of the covalent interaction cutoff distance is tuned to 2.0 Å [22-28] in the AIREBO potential. During the simulation, the overall engineering stress is tracked, which is calculated by $\sigma = F/A$. Here, $F$ is the tensile force, and $A$ is effective cross-sectional area. Specifically, the cross-sectional area is approximated as $b \times h$, with $b$ and $h$ as the width and thickness of the sample, respectively. The graphite interlayer distance 3.35 Å is taken as the thickness of the sample. Correspondingly, the engineering strain is estimated by $\varepsilon = (l - l_0)/l_0$, where $l$ and $l_0$ are the instantaneous and initial length of the sample. For comparison purpose, the effective Young's modulus is extracted from the stress-strain curve by linear fitting. The atomic stress is calculated based on the virial stress, which is expressed as [29]

$$\Pi^{\alpha\beta} = \frac{1}{\Omega}\sum_i \varpi_i \pi_i^{\alpha\beta}, \quad \pi_i^{\alpha\beta} = \frac{1}{\varpi_i}\left(-m_i v_i^\alpha v_i^\beta + \frac{1}{2}\sum_{j \neq i} F_{ij}^\alpha r_{ij}^\beta\right) \tag{1}$$

Here $\pi_i^{\alpha\beta}$ is the atomic stress associated with atom $i$. $\varpi_i$ is the effective volume of the $i$th atom and $\Omega$ is the volume of the whole system. $m_i$ and $v_i$ are the mass and velocity of the $i$th atom, respectively. $F_{ij}$ and $r_{ij}$ are the force and distance between atoms $i$ and $j$, respectively, and the indices $\alpha$ and $\beta$ denote the Cartesian components.

## 3. Results and discussions

### 3.1 Heterojunctions with graphene at the two ends

Firstly, we assess the mechanical properties of the HJs with graphene locating at the two ends of the structure. Figure 2 shows the stress-strain curves of the HJs containing different percentages of α-GY. As is seen, there are two categories of stress-strain profiles. For the sample with low α-GY percentage, the stress experiences a sharp drop after linear elastic increase, which is regarded as a brittle behaviour (though the stress does not go straight to zero), whereas, for the sample with higher α-GY content, the stress experiences another increase portion followed by a hardening period (with fluctuations), which is designated as a ductile behaviour. For both categories, the stress shows a zigzag changing profile after the abrupt stress decrease. Following the continuum mechanics, effective Young's modulus is derived from the stress-strain curves with strain up to 4%, the yield stress is defined as the maximum stress during the whole elastic tensile deformation with the corresponding strain denoted as yield strain, and the fracture stress/strain is the stress/strain when the specimen fails (see Figure 2). As expected, Young's modulus increases with the increasing percentages of graphene. Interestingly, the yield stress fluctuates in the vicinity of a certain value (~ 5.8 GPa) among all examined samples, whereas the yield strain and fracture strain exhibit a big difference. For example, the pure α-GY model shows the largest fracture strain of 37.2% and its counterpart with 90% of graphene only exhibits a yield/fracture strain of 5.3% (more than six times smaller). We should note that the estimated mechanical properties for the pure α-GY from current work diverge from that reported earlier by Zhang et al [6], which is considered as a result from the different stress calculation approach and also the boundary conditions being utilized. In this regard, we have also checked the stress-strain curve for the pure α-GY as derived from the virial stress approach, which agrees with that from Zhang et al [6]. Although the usage of different calculation approaches and boundary conditions will influence the absolute values of the interpreted mechanical properties, their scaling behaviours as focused in this work will not change. Overall, it is found that the heterojunctions with high α-GY percentage own a good ductility, i.e., possess a large fracture strain.

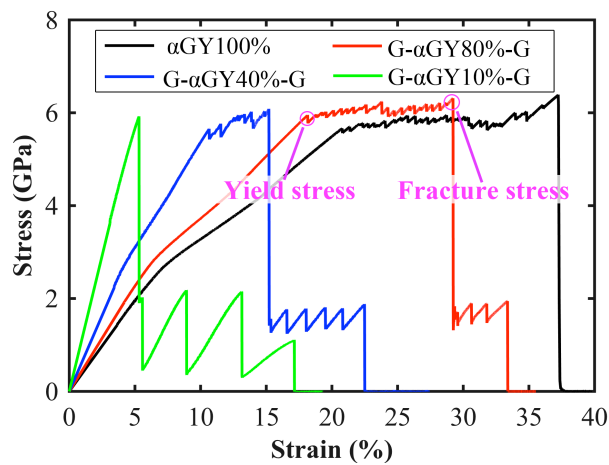

**Figure 2** Comparisons of the stress-strain curves of HJs with different percentages of α-GY locating in the middle of the structure.

To explore the underlying deformation mechanisms, we study the atomic configurations of these HJs at different strains. For the HJs with low α-GY percentage, the engineering stress drops sharply after the elastic limit, which corresponds to the brittle behaviour. For the HJs with higher percentage of α-GY, a more complex deformation processes are identified, reflecting a ductile behaviour as indicated by the stress-strain curve. Figure 3 illustrates a representative deformation process from the HJ with 40% α-GY. Generally, five deformation stages have been identified as discussed below.

At stage I, the whole structure maintains its initial configuration and only stretched bonds are observed. Continuing stretching the structure, we observe the break of acetylenic linkages, which is followed by the formation of carbon triangle rings as illustrated in Figure 3b. Such deformation phenomenon contributes to the further stress increase after the linear increasing portion, i.e., the hardening process (stage II). At the end of stage II (strain of ~ 15.2%), the connecting bonds between graphene and α-GY are extensively stretched (Figure 3c). As plotted in Figure 3c, the acetylenic linkages at the interface region experience a severer stress concentration, which triggered the interface fracture with increasing stretch (stage III). Afterwards, an interesting "tearing stages" (stage IV) is observed. As illustrated in Figures 3d and 3e, we found the tearing of the hexagonal carbon rings at the connecting edge of the graphene, which well explains the stress increase and decrease event (zigzag-shaped) as shown in Figure 3a. Accompanying with the tearing process, the two initial acetylenic linkages (see Figure 3d) are extended, forming two long monoatomic chains as shown in Figure 3d. Before the final failure of the structure (stage V), another monoatomic chain is observed (Figure 3f). In all, it is found that the considered HJs with high α-GY content ($\geq$ 40%) located in the middle of the structure uniformly exhibit five deformation stages. Particularly, the duration of the hardening process (stage II) increases with the percentage of α-GY due to the fact that more acetylenic linkages are introduced to the HJs with more α-GY.

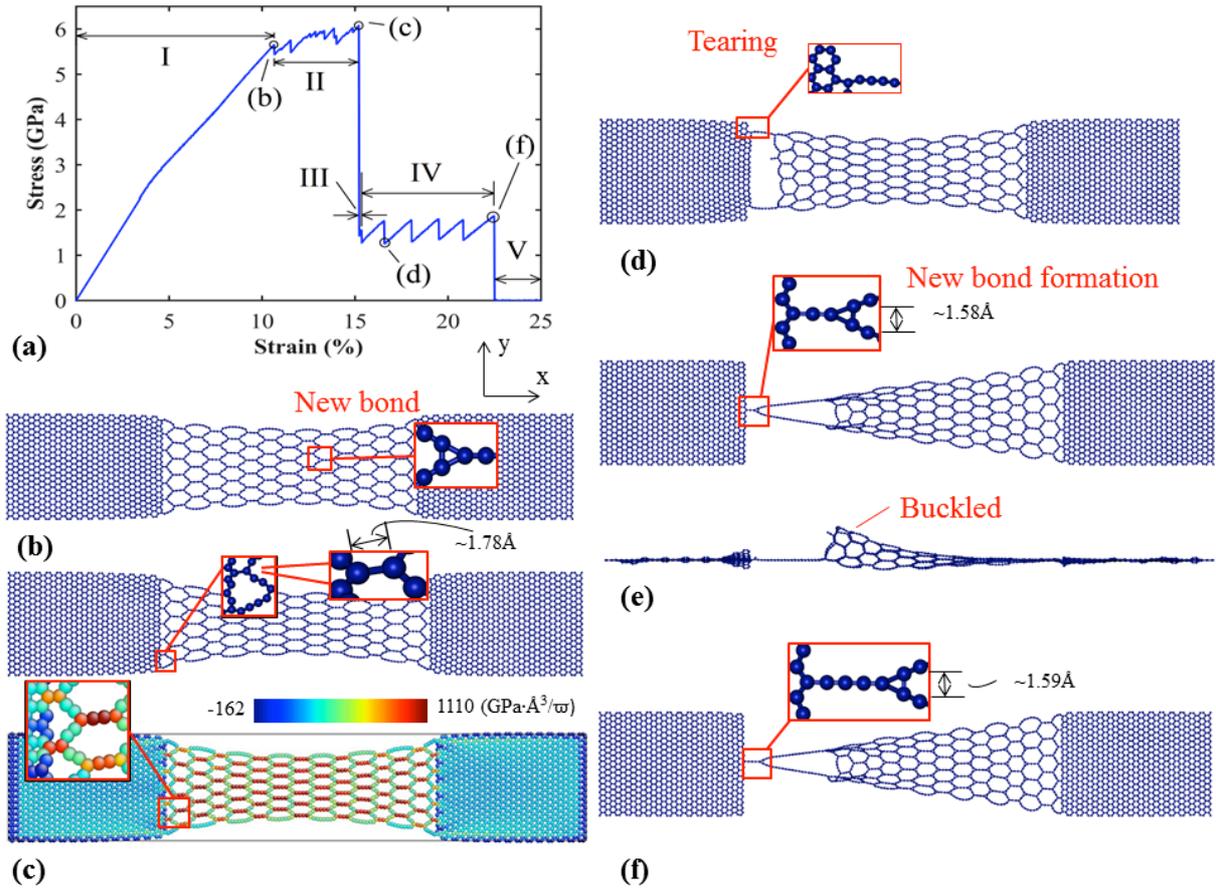

**Figure 3** (a) A representative five deformation stages from the HJ with 40% α-GY locating in the middle of the structure, including: stage I (elastic deformation), stage II (hardening), stage III (interface fracture), stage IV (tearing), stage V (finial failure). Selected atomic configurations reveal these five deformation phases of at the strain of: (b) 10.54%, (c) 15.20%, (d) 16.59%, (e) 22.49%, and (f) 22.50%. To note that the atomic configurations have been scaled to a similar size, which do not reflect the actual length of the stretched structure.

We then consider another four groups of HJs with different types of graphene allotropes, respectively, i.e., β-GY, γ-GY, 6612-GY and GDY as schematically illustrated in Figure 1. Basically, a similar stress-strain pattern/tendency is found in these four groups. The representative results from the HJs with γ-GY are plotted in Figure 4 (see Supporting Information for the results from the HJs with 6612-GY). In general, by varying the percentages of the graphene allotropes from 10 to 100%, continuous increase pattern in yield strain is observed which is similar to that of the HJs with α-GY, though the increment is much smaller. Meanwhile, the effective Young's modulus is found to decrease with increasing percentage of graphene allotropes. Also, the HJs with graphene allotropes possess a similar yield stress, but smaller than that of the pure graphene allotrope structures. For example, the average yield stress for the HJs with 60%, 40% and 10% of γ-GY is ~ 6.7 GPa, and the pure γ-GY structure shows a yield stress around 7.8 GPa. Most interestingly, no hardening process is observed and the stress of the four groups HJs is observed to experience a sharp reduction after passing the threshold value (i.e., the yield stress), signifying a brittle behaviour.

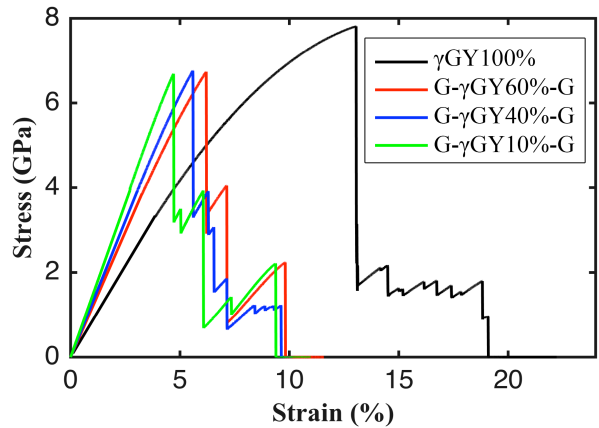

**Figure 4** A representative stress-strain curves from the HJs with different percentages of γ-GY locating in the middle of the strcuture.

In line with the similar stress-strain profiles, the HJs with different percentages of β-GY, γ-GY, 6612-GY and GDY share similar brittle deformation process but differ from that their counterpart with α-GY. That is, the structure experiences elastic deformation before yielding (with the whole structure unchanged), and then monoatomic carbon rings are formed at the interface region after yielding. Figure 5 demonstrates a representative deformation process from the HJ with 60% of γ-GY. Corresponding to the stress-strain curve in Figure 5a, the HJ is under elastic deformation before reaching the yield strain and retains its initial structure despite the stretched bonds. As plotted in Figure 5b, the atomic configuration shows that the acetylenic linkages near the interface region experience higher stress during stretch, indicating a local stress concentration. At the onset of yielding, bond breaking is initiated almost simultaneously at those stress concentration locations around the interface (see Figure 5c and 5d). More importantly, the local stress increase events in Figure 5a can be well explained according to the corresponding atomic configurations. As plotted in Figure 5e, at the strain of ~ 6.2%, the HJ exhibits a symmetrical structure along the width direction and the left and right regions are connected by four carbon bonds. Stretching of these four bonds is observed to induce a stress increase event, and the breaking of the two outside bonds (Figure 5f) is found to lead to the stress sharp decrease. After the fracture, a closed monoatomic carbon ring is retained (Figure 5g).

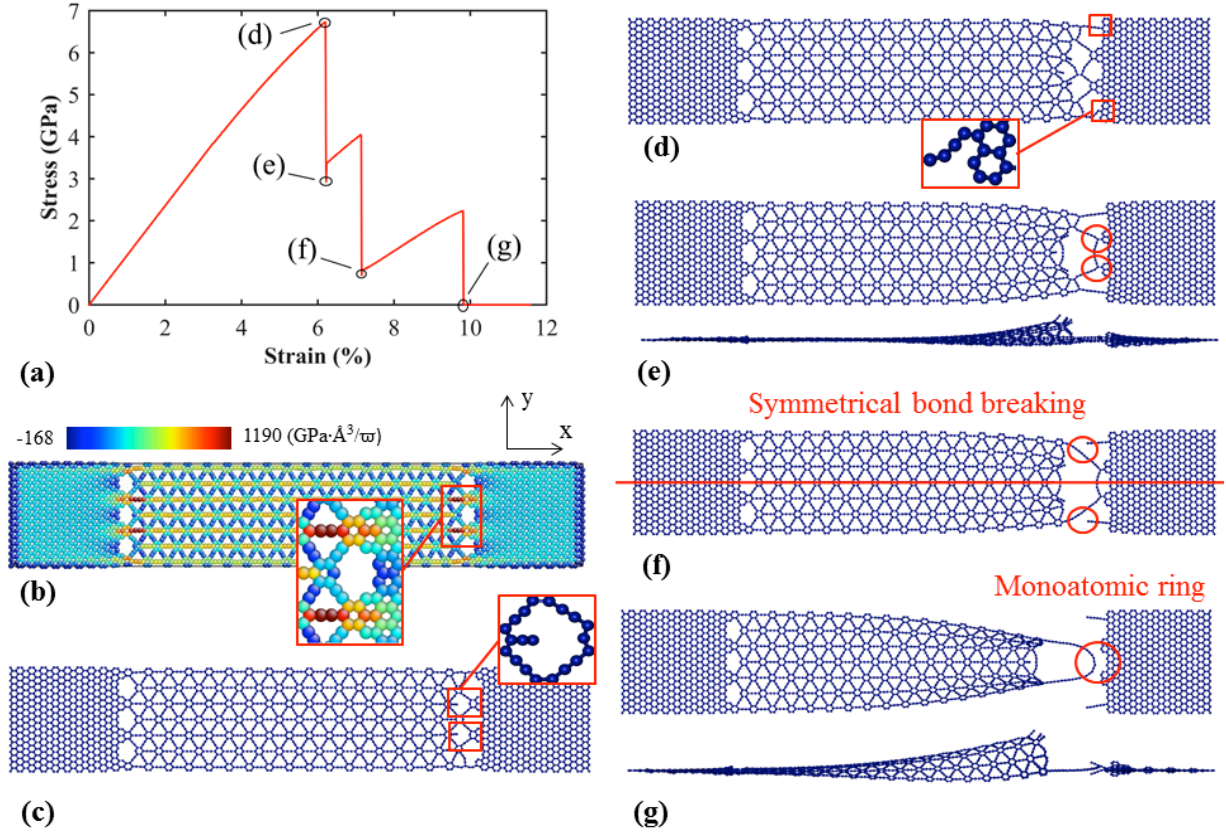

**Figure 5** A representative deformation process from the HJ with 60% of γ-GY locating in the middle of the structure: (a) Stress-strain curve; Atomic configurations of the HJs at the strain of: (b) 6.16%, (c) 6.19%, (d) 6.20%, (e) 6.22%, (f) 7.14%, and (g) 9.82%. To note that the atomic configurations have been scaled to a similar size, which do not reflect the actual length of the stretched structure.

### 3.2 Heterojunctions with graphene in the middle

Above discussions suggest that the tensile mechanical properties of HJs can be tailored by changing either the type or the percentage of graphene allotropes. In this section, we assess how the HJs will behave under tensile while the graphene section is in the middle of the structure. Again, we start with the HJ with different percentages of α-GY. Consistent with the observations in Sec. 3.1, the increasing percentages of graphene is found to result in a continuous reduction to the yield/fracture strain of the structure, and there are two categories of deformation behaviours including a brittle behaviour for HJ with high percentage of graphene and ductile behaviour for HJ with low percentage of graphene. Comparing with the results in Figure 2, the stress-strain curves shows similar pattern, however shorter hardening period is observed comparing with that of the HJ with same amount of α-GY in the middle of the structure. For example, no hardening process is occurred for the HJ with 40% of α-GY at the two ends according to the stress-strain curve in Figure 6 (red curve), which however is observed from its counterpart with the α-GY in the middle of the structure (Figure 2, blue curve). Such phenomenon is partially originated from the fact that part of the α-GY (at the two ends of the structure) has been frozen as boundaries, and thus less α-GY is involved in the tensile deformation.

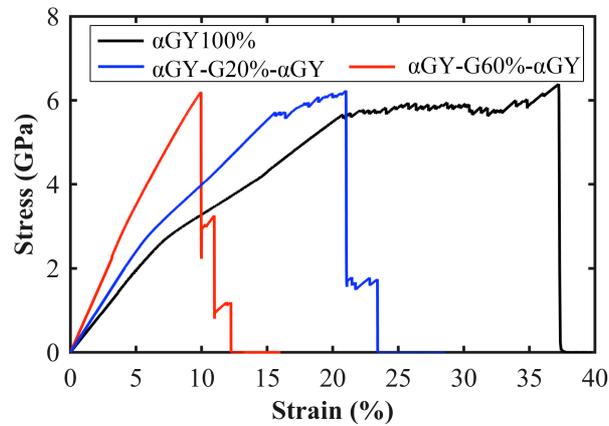

**Figure 6** Comparisons of the stress-strain curves of HJs with different percentages of α-GY locating at the two ends of the structure.

To explore the tensile deformation with no hardening process, we investigate the atomic configurations of the HJ, which contains 40% α-GY locating at the two ends. As illustrated in Figure 7a, the acetylenic linkages in the HJs along the length direction absorb the majority of the strain energy, which induce local stress concentrations. Unlike the HJs with low graphene percentage, no new bond formation is observed after passing the yield strain (Figure 7b). Instead, we found the breaking of the acetylenic linkages adjacent to the connecting interface, which agrees with the brittle behaviour as indicated by the stress-strain curve. After yielding, further stretching of the structure leads to the formation of monoatomic rings in the deformation region (Figures 7c and 7d). Before failure, we also notice a new bond formation which leads to a triple carbon rings as shown in Figure 7e.

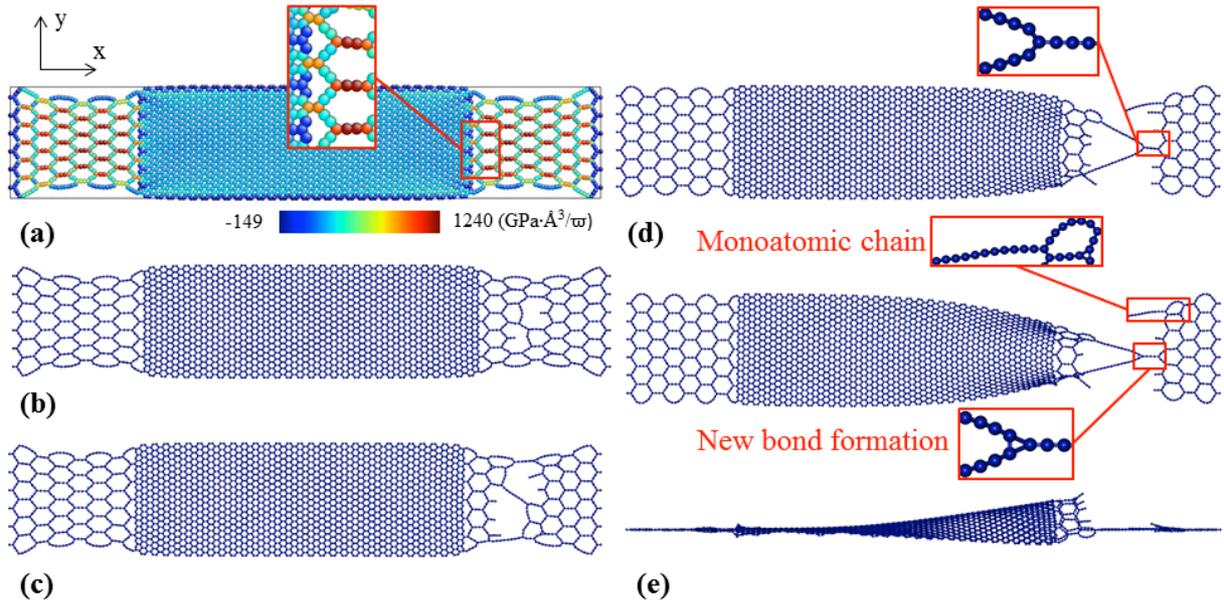

**Figure 7** Tensile deformation of the HJ with 40% of α-GY at the two ends of the structure. Atomic configurations of the structure at the strain of: (a) 9.95%, (b) 10.01%, (c) 10.98%, (d) 11.80%, and (e) 12.27%. To note that the atomic configurations have been scaled to a similar size, which do not reflect the actual length of the stretched structure.

Investigations are also extended to the HJ with other different graphene allotropes (β-GY, γ-GY, 6612-GY and GDY) locating at the two ends of the structure. Similar as previous results, these graphene allotropes are found to share a similar brittle behaviour. Figure 8 shows the representative stress-strain curves from the HJ with different percentages of γ-GY (the results from the HJs with 6612-GY are also included in the Supporting Information). Basically, with the increase of the graphene percentage, the yield strain decreases, the effective Young's modulus increases, and the yield stress fluctuates around a certain value (and it is much smaller than that of the pure γ-GY structure).

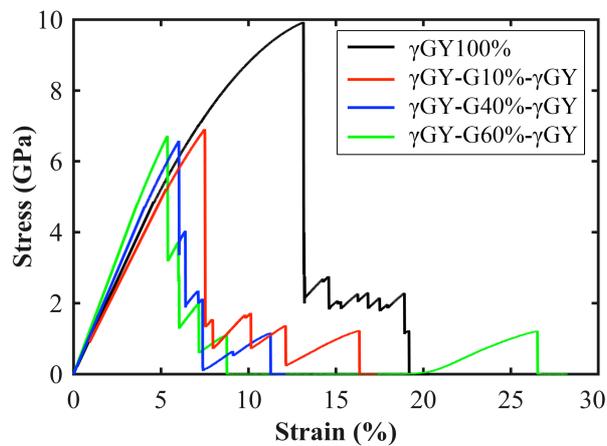

**Figure 8** Comparison of the stress-strain curves for the HJs with different percentages of γ-GY at the two ends of the structure.

The deformation mechanism and stress distribution pattern are also similar as observed from the HJ with γ-GY at the middle of the structure, as is seen from the atomic configurations of the HJ with 40% of γ-GY in Figure 9. Specifically, the HJ retains its initial structure before yielding and the acetylenic linkags along the tensile direction experience higher stress (induce local stress concentration, Figure 9b). After yielding, each bond breaking results in a sharp stress decrease event as highlighted in Figure 9a. It is worthy to mention that the stress of the HJ with 40% of γ-GY appears nearly zero from the strain ~ 8.8% to ~ 20% and then resumes to around 1 GPa. Such interesting phenomenon is due to the relaxation of the whole structure. As presented in Figure 9f, a long monoatomic ring is formed at the strain of 8.8%, which allows the previously strained HJ relaxing to the nearly zero stress state. Further elongation will stretch the monoatomic ring, and thus leads to the stress increase event. The atomic configurations of the HJ with 90% of 6612-GY are also included in the Supporting Information for comparison purpose.

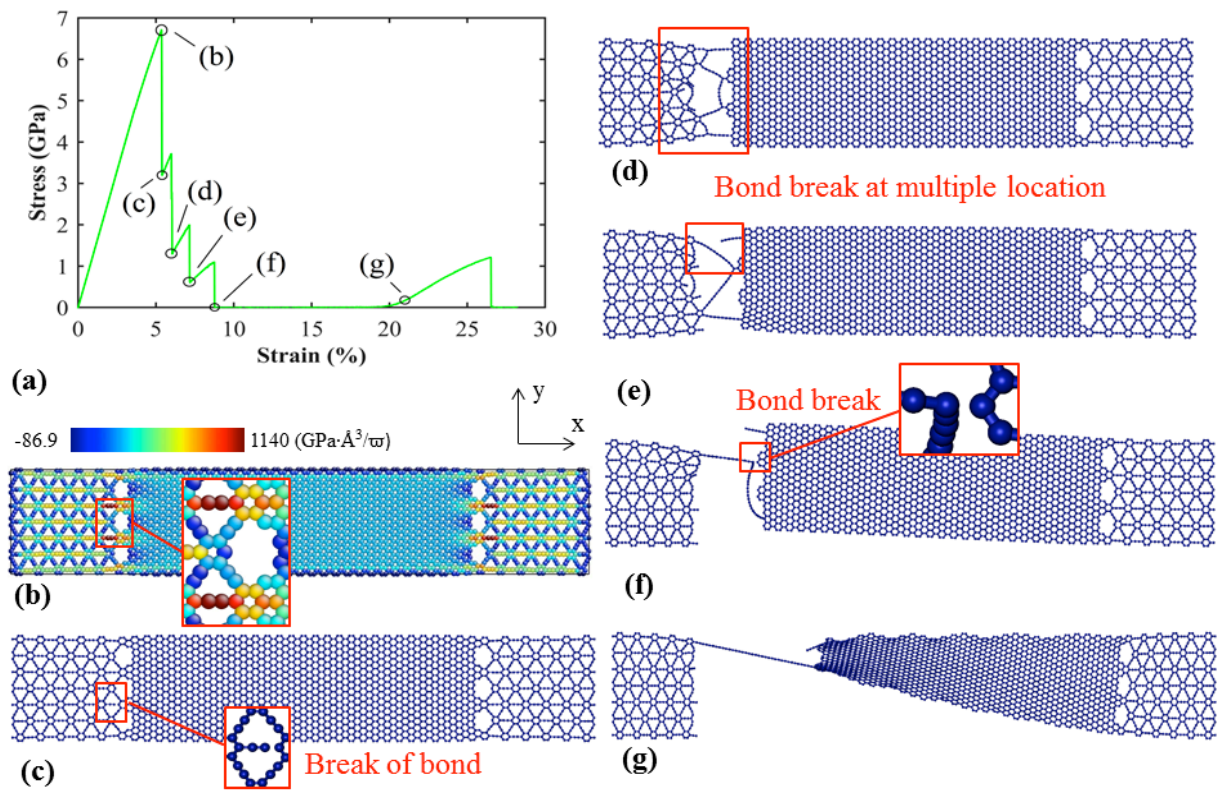

**Figure 9** A representative tensile deformation process from the HJ with 40% of γ-GY at the two ends of the structure: (a) Stress-strain curve; Atomic configurations at strain of: (b) 5.35, (c) 5.41%, (d) 6.07%, (e) 7.19%, (f) 8.78%, and (g) 20.98%. To note that the atomic configurations have been scaled to a similar size, which do not reflect the actual length of the stretched structure.

## 3.3 Discussions

Evidently, above results indicates that the tensile properties of the HJ can be tuned by varying the type, percentage or location of the constituent graphene allotropes. In this section, we compare the yield strain/stress and Young's modulus among all studied HJs. It is found that the location of graphene does not have a significant impact on the mechanical properties of the HJ, i.e., the yield strain/stress and Young's modulus exhibit a similar relationship with the percentage of graphene when the graphene locations changed. The results from the HJs with graphene locating at the two ends of the structure are summarised in Figures 10a, b and c (see Supplementary Information for the summarized results from the HJs with graphene locating in the middle). Generally, the yield strain decreases with the graphene percentage (Figure 10a), the yield stress appears irrelevant with the graphene percentage (Figure 10b), and the estimated Young's modulus increases with the graphene percentage (Figure 10c). Specifically, the HJs with α-GY usually exhibit higher yield strain comparing with other counterparts (Figure 10a), and the HJs with 6612-GY possess higher yield stress.

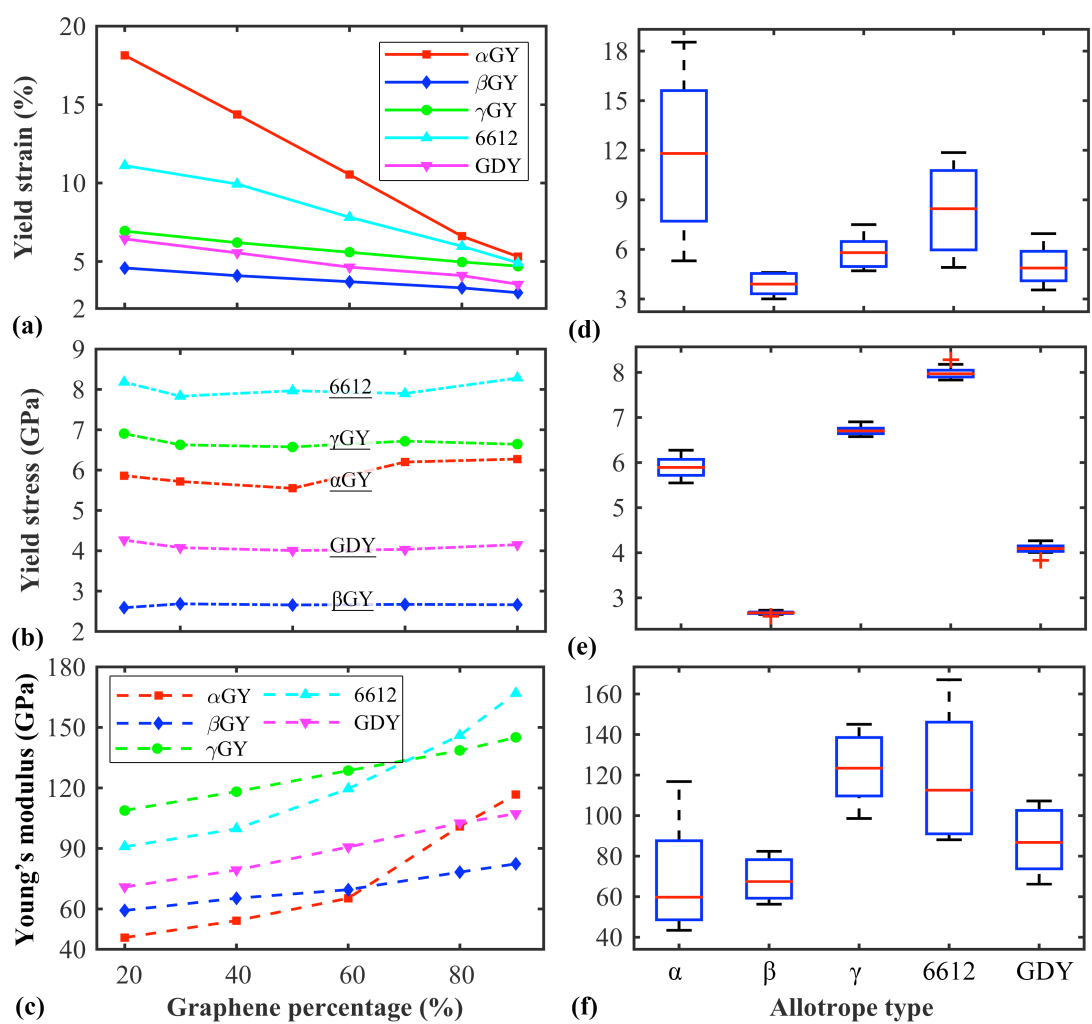

**Figure 10** (a) The yield strain, (b) the yield stress, and (c) the effective Young's modulus, as a function of the graphene percentage for HJs with graphene locating at the two ends of the structure; Overall comparisons of the mechanical properties of HJs with graphene locating in the middle and at the two ends of the structures in boxplot: (d) the yield strain, (e) the yield stress, and (f) the effective Young's modulus.

We should highlight that the HJ with higher graphene percentage tends to earlier yielding (Figure 10) despite the fact that graphene owns superior mechanical properties comparing to those graphene allotropes [6, 30]. Such surprising results can be explained according to the continue mechanics by considering a simple heterojunction nanoribbon as schematically shown in Figure 11. Supposing that the graphene section has a length of $L_G$ and Young's modulus of $E_G$, and the allotrope section has a length of $L_A$ and Young's modulus of $E_A$, then the force in each section should equal to each other under certain tension, i.e.,

$$E_A \frac{\Delta L_A}{L_A} A = E_G \frac{\Delta L_G}{L_G} A \qquad (2)$$

Here, a uniform cross-sectional area $A$ is assumed for the whole sample. $\Delta L_A$ and $\Delta L_G$ represent the elongation in the allotrope and graphene regions, respectively. Therefore, considering that the sample length $L = L_A + L_G$ and the overall elongation $\Delta L = \Delta L_A + \Delta L_G$, the strain $\varepsilon_A$ in the allotrope section can be related to its initial length by

$$L + \left(\frac{E_G}{E_A} - 1\right) L_A = \frac{E_G}{E_A} \Delta L \frac{1}{\varepsilon_A} \qquad (3)$$

Given that for HJ with different percentages of graphene allotropes, the sample length $L$ and Young's modulus ratio $E_G / E_A$ are constants, then at the same strain or same overall elongation the strain $\varepsilon_A$ is inversely related to the initial length $L_A$. That is, $\varepsilon_A = 1/(a + bL_A)$, with $a$ and $b$ as constants. In other words, decrease the percentage of graphene allotrope (i.e., increase the graphene content) will induce higher strain in the allotrope section, which therefore leads to earlier yielding (as the yield stress of those graphene allotropes are much lower than that of graphene [6, 30]).

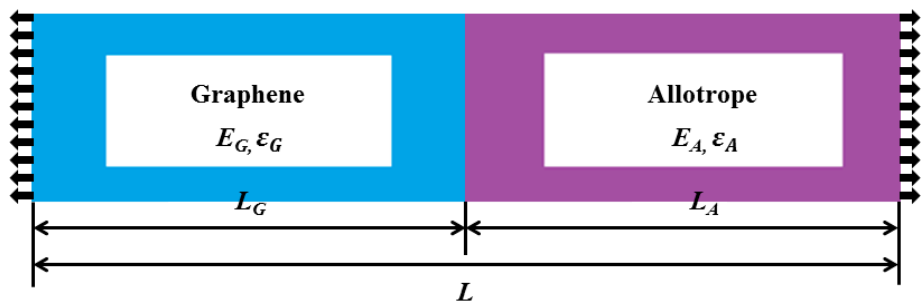

**Figure 11** Schematic view of a continuum model for the heterojunction.

In all, the various mechanical properties of the HJ as resulted from the presence of different graphene allotropes suggests the tunability of its structure and thus mechanical performance to facilitate various application purpose. As compared in Figure 10d, e and f (graphene percentage ranging from 10% to 90%), the HJ with α-GY possesses the widest tuning ranging for yield strain (from 5.3% to 18.5%) and Young's modulus (from 43.4 GPa to 116.8 GPa). Followed by the HJ with 6612-GY, which also shows a wide adjusting range for the yield strain (from 4.9% to 11.9%) and maintains a relatively higher Young's modulus (from 88 to 167 GPa). It is noticed that the HJ with β-GY is usually show the smallest yield strain/stress and lowest Young's modulus, which leaves seldom spaces for the tensile properties engineering purpose. Besides, we should note that the presence of α-GY in the HJ structure also results in a hardening process, which greatly extends the ductility of the heterojunctions, suggesting a great potential for the applications as flexible/stretchable nano-electronics.

## 4 Conclusions

Basing on a series of *in silico* studies, we assessed the tensile properties of 2D heterojunctions (HJs) made from graphene and different graphene allotropes (including α-, β-, γ- and 6612-GY, and GDY). It is found that all HJs exhibit a brittle behaviour except the one with α-GY, which however shows a hardening process due to the formation of triple carbon rings. The structure usually yields at the interface between graphene and graphene allotropes, and monoatomic carbon rings are usually formed after yielding. The similar mechanical properties obtained from the HJs with graphene either located in the middle or at the two ends suggest that the location of graphene allotropes exerts insignificant impacts. In the other hand, different Young's modulus, yield stress, and yield strain are found for the HJ with different types and percentages of graphene allotropes. Generally, it is found that the yield strain decreases with the graphene percentage and the yield stress behaves irrelevant with the graphene percentage. In comparison, the estimated Young's modulus increases with the graphene percentage. This study provides a fundamental understanding of the tensile properties of the heterojunctions constructed from graphene and graphene allotrope, which should guide the design and engineering of their mechanical properties, and eventually benefit their future applications in nanoscale devices, such as flexible/stretchable electronics.

## Acknowledgments

Supports from the ARC Discovery Project (DP130102120), the High Performance Computer resources provided by the Queensland University of Technology are gratefully acknowledged.

## Supporting Information

Supporting information is available for the simulation results from the HJs containing 6612-GY, and the summarised mechanical properties of the HJs with graphene locating in the middle of the structure.